\documentstyle[aps,preprint,amstex]{revtex}

\def\squote{}
\def\quote#1#2#3#4{\squote {#1,\ {\sl#2}\ {\bf#3}, #4}.\par} 
\def\qquote#1#2#3#4{\squote {#1,\ {\sl#2}\ {\bf#3}, #4};}

\def\prl{{\sl Phys. Rev. Lett.}\ }

\def\pr {{\sl Phys. Rev.}\ }
\def\ksi{\xi}
\def\mr{magnetoresistance }

\def\dos{density of states }

\def\e{\epsilon}

\def\vrh{variable-range hopping\ }
\def\so{spin-orbit\ }
\begin{document}
\title{Spin-Orbit Assisted Variable-Range Hopping in Strong Magnetic
Fields}
\bigskip
\author{\large Yigal Meir}
\address{Physics Department, Ben Gurion University, Beer Sheva 84105, ISRAEL}
\author{\large Boris L. Altshuler}
\address{NEC Research Institute, 4 Independence Way, Princeton, NJ 08540}
\maketitle
\begin{abstract}
It is shown that in the presence of strong magnetic fields,
spin-orbit scattering causes a sharp increase in
the effective density of states in the variable-range hopping regime when
temperature decreases. This effect leads to an exponential enhancement of the
conductance above its value without spin-orbit scattering.  Thus an
experimental study of the hopping conductivity 
in a fixed, large magnetic field,
 is a sensitive tool to explore the spin-orbit scattering parameters in
the strongly localized regime.
\end{abstract}
\pacs{72.20.My, 72.20.-i, 73.50.Jt}
While the effects of \so scattering in the weakly localized regime are well
understood \cite{bergmann},  much less is known on how spin-orbit scattering
affects the transport in the strongly localized regime. Several different
effects due to \so scattering have been suggested
 \cite{mwea,shapir,sanquer,kardar,eto},  with different,  and sometimes contradicting predictions. All
 of these works rely on the measurement of the magnetoresistance, just as
 in the weakly localized regime,  in order to explore the \so effects. However, 
 the interplay of the different mechanisms leading to \mr in the strongly
 localized regime ambiguates the experimental results,  leading to indefinite
conclusions.
 
 In this work we propose a different approach which will allow determination 
 of the \so scattering parameters without relying on \mr  measurements. In
 particular,  we show that in the presence of strong magnetic field,  \so
 scattering leads to a temperature dependence of the effective density of
  states,  $\rho$,  in the Mott variable-range hopping law, 
$R \sim \exp\{(T_0/T)^{1/(d+1)}\}$, with
 $T_0 \sim 1/\rho\ksi^d$, 
and $\ksi$ the localization length. Thus \so scattering will lead to 
 an exponential change in the resistance,  with a crossover temperature
 determined by the spin-orbit scattering and by the magnetic field.
 
 The sensitivity of the \vrh resistance to \so scattering stems from an effect, 
 first pointed out by Kamimura and coworkers \cite{kamimura}. At zero magnetic
 field each impurity state can be either unoccupied,  singly occupied
 or doubly occupied. Hopping processes can occur from a singly or doubly occupied
 state to an unoccupied or singly occupied state. However, 
a strong magnetic field
 $\left(g\mu H > kT \{(T_0/T)^{1/(d+1)}\right)$
\cite{glazman,myvrh}  polarizes all singly occupied states,  thus 
 blocking all singly occupied to singly occupied hopping processes (and by
 detailed balance,  also all doubly occupied to unoccupied ones),  leading
 to an effectively reduced density of states,  and an exponentially enhanced
 magnetoresistance. Such a strong positive   magnetoresistance,  which saturates
 at high fields,  has indeed been observed experimentally 
\cite{jiang,frydman,essaleh}.
 
 Here we demonstrate that in the presence of \so scattering,  the
high-field
 effective \dos increases at low temperatures to its zero-field value. The
 calculation involves two steps: (a) calculation of the probability that
  the electron can spin-flip during its hop,  due to \so scattering,  and
  (b) calculation of the contribution of the spin-flip hops to the resistance.
  
 {\sl The probability of a spin-flip hop}. The hopping probability is 
 proportional to the overlap of the wavefunction the electron hops from with
 the one the electron hops onto, squared. As mentioned above,  
 in strong magnetic fields, 
 and in the absence of \so scattering,  hopping from a singly occupied state
 to a singly occupied state involves two wavefunctions with 
 opposite spins and consequently zero overlap. In the presence of \so scattering, 
 each wavefunction will acquire a component in the opposite spin direction.
 Assuming a short range \so scattering,  $V_{so}(r) = U_{so} \delta (r-r_i)$, 
 where $r_i$ is the position of the scatterer and  $U_{so}$ its strength,  the
 amplitude of the opposite-spin component, say $A_\downarrow$, due
 to that single scatterer, 
  is given by first order perturbation
 theory, 
 \begin{equation}
 A_\downarrow = {U_{so} \over {g\mu H}}\ {1\over r_i^d}\ e^{-2r_i/\ksi}
  . 
 \label{perturbation} 
 \end{equation}
 Averaging Eq.(\ref{perturbation}) over all possible positions of
the \so scatterers,  we find
 \begin{equation}
 A_\downarrow = {U_{so}\,  n_{so}\over {g\mu H}} \equiv {H_{so}\over H}    ,  
 \label{average} 
 \end{equation}
 where $n_{so}$ is the density of \so scatterers,  and where trivial numerical
 factors have been omitted. Note that the dominating \so scattering occurs
 within the localization length,  and so we expect $H_{so}$ to be similar
 to its weak-localization value \cite{bergmann}. The spin-flip hopping
 probability,  $P_{so}$,  will thus be proportional to $(H_{so}/H)^2$.
 
 {\sl The calculation of the resistance}. We follow here the approach
 of Ambegaokar,  Halperin and Langer \cite{ambegaokar}. According to
 Miller and Abrahams \cite{miller}  the resistance can be calculated by
 solving the equivalent random-resistor network,  where each pair of
 impurity states is connected by a classical resistor, 
  $R=R_0 e^{\Delta \epsilon/kT + 2r/\ksi}$, 
   where $\Delta \e$ is the difference in energy  between the states,   $r$
   is the distance between the impurities,  and $R_0$ is some microscopic
   resistance. In the following all resistances will be in units of $R_0$. 
   Ambegaokar et al. \cite{ambegaokar}
   suggested that due to the exponential spread in the values of the
   resistors, the resistance of the network will be determined
   by the lowest resistance,  $R$,  such that the network composed of
   all resistors with resistances smaller than  $R$, percolates. Since
   all such resistors have to obey $|\Delta \epsilon| < kT \log R$,  and
   $r < (\ksi/2) \log R$,  the percolation criterion takes the form
\begin{equation}
z_d = 2 \rho_0  \ kT \log R \  (\ksi \log R / 2)^d , 
\label{pc}
\end{equation}
with $z_d$ the critical density in $d$ dimensions ($z_2\simeq6.9$ and 
$z_3\simeq5.3$ \cite{skal}),  leading to the Mott hopping law
\begin{equation}
(\log R)^{d+1} = {2^{d-1}z_d\over{kT \rho_0 \ksi^d}} \equiv {T_0\over T} .
\label{percolation}
\end{equation}

As mentioned above, in strong magnetic field and without \so scattering,
the network separates into two subnetworks (denoted $A$ and $B$ in Fig. 1). 
An electron can hop from a singly
occupied site onto an unocuppied one (type $A$),
 which then becomes singly occupied,
and then hops onto an unoccupied site, and so forth. Or the electron can hop
from doubly occupied sites to singly occupied ones (type $B$). 
As can be seen from Fig.1, 
the density of the type $A$ impurities is determined by the density
of states around the Fermi energy, while that of  
  type $B$ impurities is determined by the density
of states at energy $U$ away from the Fermi energy, where $U$ is the 
intra-impurity, Hubbard-like
repulsion between the electrons. If the density of states is constant
on an energy scale $U$, then the effect of the magnetic field is to 
replace $\rho_0$ in Eq.(\ref{percolation}) by $\rho_0/2$,  leading
to the doubling of $T_0$ and 
an exponentially enhanced resistance. 

In the presence of \so scattering impurity states
with opposite spin are also connected, with resistance 
$R=e^{\Delta \epsilon/kT + 2r/\ksi}/P_{so} $. 
This will lead,  in the case of constant density of states,  to a modified
percolation criterion, 
\begin{equation}
(\log R)^{d+1} + [\log (R P_{so})]^{d+1} = {2T_0\over T}   .
\label{sopercolation}
\end{equation}
As long as $R < 1/P_{so}$,  the solution of (\ref{percolation}) will yield
a smaller resistance than the solution of (\ref{sopercolation}). Thus,  at
high temperatures spin-orbit scattering will not play any role,  as the
resistors involving spin-flips will not participate in the percolating
cluster. These will start to contribute as the temperature is lowered
so that $R > 1/P_{so}$,  and at small enough temperature 
Eq.(\ref{sopercolation}) has a solution similar to (\ref{percolation}),  
with $\rho_0/2$ replaced by $\rho_0$,  namely all types of 
hopping processes contribute
to the conductance. Thus we expect the effective $T_0$ to be temperature
dependent,  with an effective \dos $\rho_0/2$ at high 
temperatures and saturating at a value corresponding to an effective 
\dos $\rho_0$
at low temperatures. The crossover temperature is determined by
$R = 1/P_{so}$,  so it depends (see Eq.(\ref{average})) on the spin-orbit
scattering and on magnetic field. Eqs.(\ref{percolation}) and 
(\ref{sopercolation}) can  be solved exactly in 2 and 3 dimensions, 
and  one finds that
 $T_0^{\bf eff}(T) \equiv \left[ d(\log R)/d(1/T^{1/d+1})\right]^{d+1}$, 
 is given by
\begin{equation}
{{T_0^{\bf eff}}\over T_0} = 
\begin{cases}
{\displaystyle 
 8{{\left[x^{2/3}+(4+h_2)^{2/3}\right]^3}\over {h_2^3(4+h_2)}}}&\ \ \ \ \ \  2d\\
 & \\
{\displaystyle 
{8\over{(h_3-3\sqrt{x}/2\sqrt{2})^2h_3^2}}}
 & \ \ \ \ \ \  3d  
\end{cases}
\end{equation}
 with $x=|\log P_{so}|^{d+1}T/T_0$,  $h_2=\sqrt{16+x^2}$
  and $h_3=\sqrt{x+2}$.
 These expressions are valid for $\log R>|\log P_{so}|$. For
 higher temperatures $T_0^{\bf eff}$ is given by its high temperature value, 
 $2T_0$.

When the density of states is not constant,  one has first to calculate
the probabilities, $p_A$ and $p_B$,  of an impurity of type $A$ or $B$, 
respectively, to belong to the 
percolating cluster. To see that these probabilities are not given by
the relative densities,  $\rho_A$ and $\rho_B$,  respectively,  one can
look at the strong field limit. In this case,  since the two subnetworks
do not communicate, the percolating
network will consist only of one type of impurities,  that with the
larger density. So if $\rho_A>\rho_B$,  then $p_A=1$ and $p_B=0$ independent
of the values of $\rho_A$ and $\rho_B$.

In order to derive an equation for $p_A$,  we realize that in order for
an impurity to belong to the infinite cluster,  it has to be connected
to another impurity on that cluster. This latter impurity can be either
of type $A$ or type $B$,  and the resistor between these two impurities 
has to satisfy the condition (\ref{pc}). Thus we find the self-consistent
equation, 
\begin{equation}
{p_A\over p_B} \ = \ {\rho_A \over \rho_B}\ \  
{{p_A (\log R)^{d+1} + p_B [\log (R P_{so})]^{d+1}}
\over {p_A [\log (R P_{so})]^{d+1} + p_B (\log R)^{d+1}}} .
\label{pab}
\end{equation}
Eq.(\ref{sopercolation}) for the resistance becomes
\begin{equation}
1 = {T\over {T_0}}\left\{(p_A\rho_A+p_B\rho_B)(\log R)^{d+1}
+ (p_A\rho_B+p_B\rho_A)  [\log (R P_{so})]^{2d+2} \right\}  .
\label{Rab}
\end{equation}
Combining Eqs.(\ref{pab}) and (\ref{Rab}),  we arrive at the final 
equation for the resistance, 
\begin{equation}
1 = {T\over {2T_0}}\left\{ (\log R)^{d+1} + \sqrt{(\log R)^{2d+2}
(\rho_A-\rho_B)^2 + 4 \rho_A \rho_B [\log (R P_{so})]^{2d+2} } \right\} .
\label{Rso}
\end{equation}
In Fig.~2 we plot  $T_0^{\bf eff}/T_0$,  for $\rho_A=\rho_B$,  and
for $\rho_A=2\rho_B$. Indeed, 
 we see that the effective $T_0$ reduces from its high-temperature
value ($T_0/\max\{\rho_A, \rho_B\}$) to $T_0$,  due to the increase in the
effective density of states. The transition starts to occur when $\log R =
1/\log P_{so}$, 
  and approaches smoothly its low temperature limit.

To conclude,  we have made detailed predictions how to determine the 
spin-orbit
scattering parameters in the strongly localized regime. The suggested
 measurements are performed in a constant magnetic field,  and avoid the 
 complications arising from the various contributions to the resistance
 as the magnetic field changes. We predict that the
effective density of states,  at large magnetic fields,  will  start to
deviate from its high temperature value at temperature given by
$T=T_0/|\log P_{so}|^{d+1}$,  where $P_{so}$ can be controlled both by the
strength of spin-orbit scatterers and their density,  and by the magnetic 
field. We predict that the effective density of states will approach at low
temperatures a value, about twice its high temperature value,  depending
on the uniformity of the density if states. These changes
in the effective density of states can be easily probed by the changes 
in the exponent in the Mott variable-range resistance.
\newpage
\leftline{\sl Figure Captions}
1. Schematic picture of the impurity states in strong magnetic fields.
Due to the polarization of the singly occupied states,  only hopping 
processes (denoted by arrows) 
between impurities of type $A$,  or between impurities of type $B$, 
are allowed in the absence of spin-flip process,  leading to a reduction
in the effective density of states.\par
\bigskip
2. The effective $T_0$,  appearing in the Mott variable-range hopping
formula. $T_0^{\bf eff}$ changes from its high temperature value 
($T_0/\max\{\rho_A, \rho_B\}$) to its low temperature value, $T_0$, due
to the increasing relevance of spin-flip processes at low temperature, 
leading to an increase of the effective density of states.

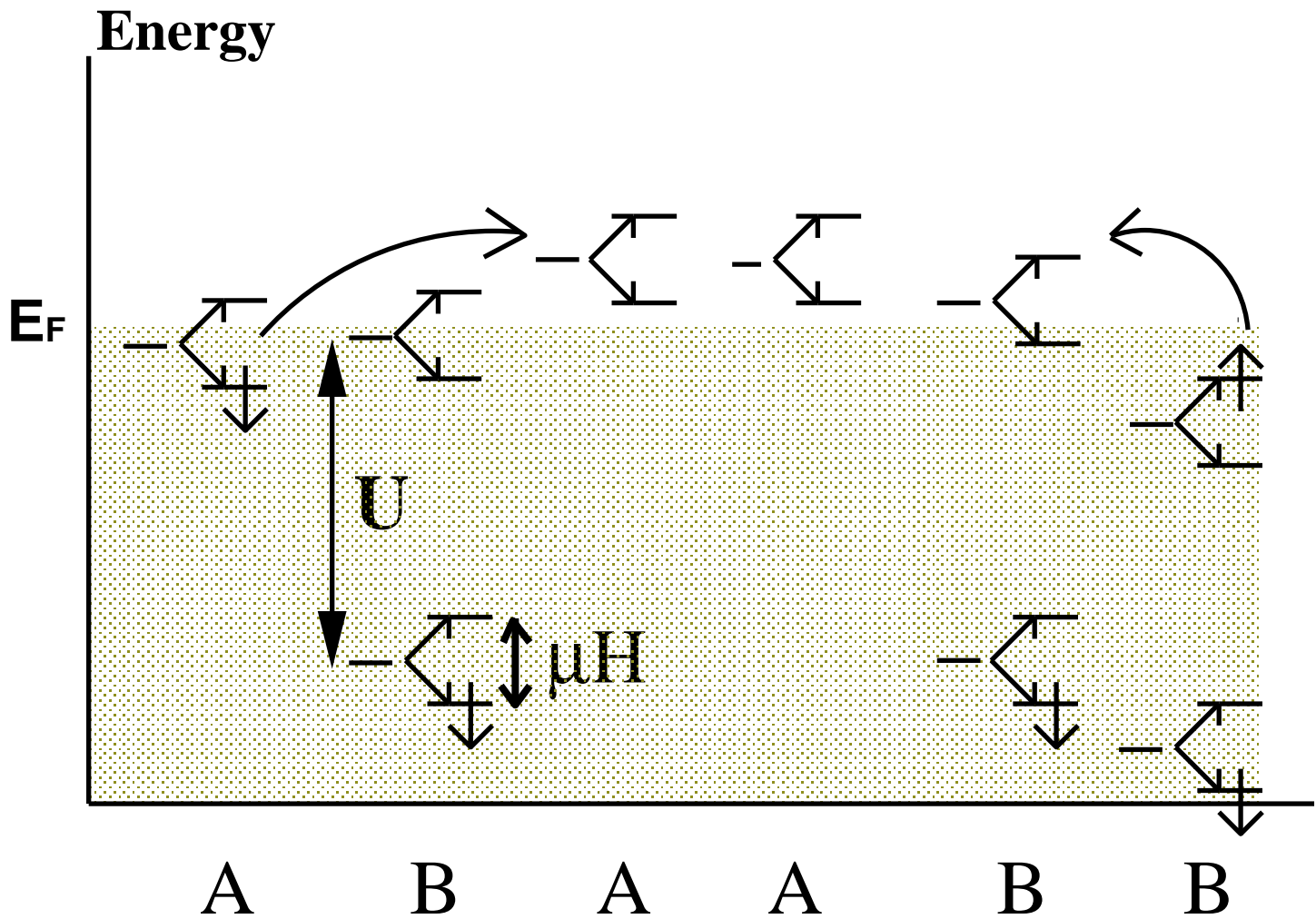
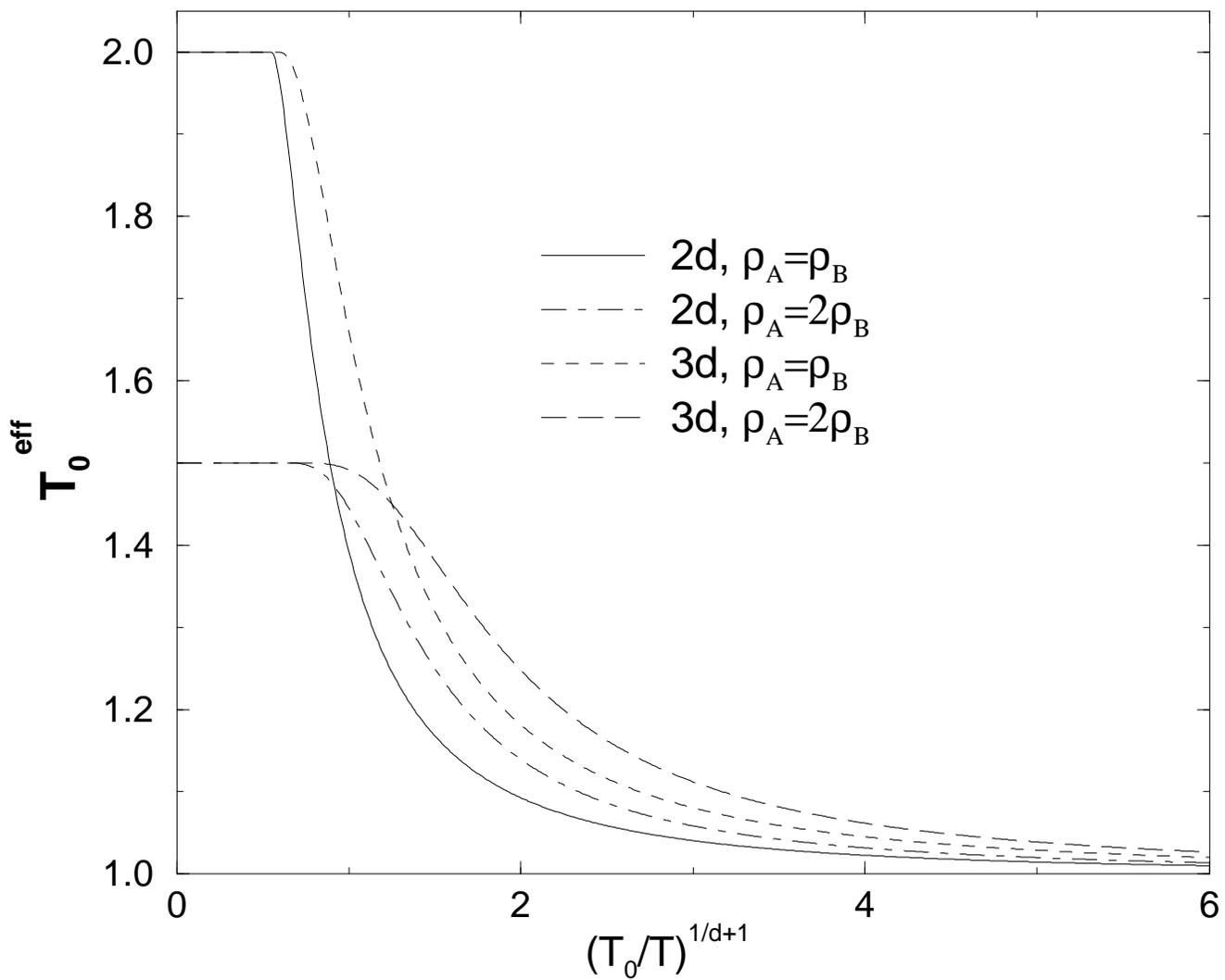
\end{document}